\documentclass[pdflatex,sn-mathphys-ay]{sn-jnl}

\usepackage[T1]{fontenc}
\usepackage[utf8]{inputenc}
\usepackage{microtype}
\usepackage{amsmath,amssymb,bm}
\usepackage{booktabs}
\usepackage{tabularx}
\usepackage{graphicx}
\usepackage{siunitx}
\usepackage{xcolor}
\let\orcidlogo\relax

\usepackage{orcidlink}
\usepackage[nameinlink,noabbrev]{cleveref}

\hypersetup{
  pdftitle={Toward Interpretable Quantum Generators for Financial Returns through Problem-Informed Circuit Design},
  pdfauthor={Huijie Guan},
  pdfsubject={Research article}
}

\graphicspath{{figures/}{../quantum_circuit_schematics/}}
\title[Problem-informed quantum generators for financial returns]{Toward Interpretable Quantum Generators for Financial Returns through Problem-Informed Circuit Design}
\author*[1]{\fnm{Huijie} \sur{Guan}\,\orcidlink{0009-0000-7348-0461}}
\email{huijie.guan@quantum-info.com}

\author[2]{\fnm{Xuanhao} \sur{Wang}}

\author[3]{\fnm{Shengbin} \sur{Wang}\,\orcidlink{0009-0001-5961-2752}}

\author[1]{\fnm{Jiale} \sur{Zhang}\,\orcidlink{0009-0006-9375-0469}}

\author[3]{\fnm{Jian} \sur{Ding}}

\author[1]{\fnm{Shuxin} \sur{Xie}}

\author[1]{\fnm{Dongdong} \sur{Li}\,\orcidlink{0000-0002-7920-0835}}

\author[3]{\fnm{Jianjian} \sur{Gao}}

\author[3]{\fnm{Hantao} \sur{Sun}}

\author*[1]{\fnm{Wenhao} \sur{Chu}\,\orcidlink{0009-0007-1395-0610}}
\email{wenhao.chu@quantum-info.com}

\affil*[1]{\orgname{QuantumCTek Co., Ltd.},
  \orgaddress{\city{Hefei}, \postcode{230088}, \country{China}}}

\affil[2]{\orgname{Guanghua Cambridge International School},
  \orgaddress{\city{Shanghai}, \postcode{201319}, \country{China}}}

\affil[3]{\orgname{China Telecom Quantum Information Technology Group Co., Ltd.},
  \orgaddress{\city{Hefei}, \postcode{230088}, \country{China}}}

\begin{document}

\abstract{
When used as function approximators, quantum circuits often lack interpretability and face a fundamental trade-off between expressivity and trainability. We propose a problem-informed framework for financial return generation that uses relational structure and symmetry-inspired parameter sharing to guide the construction of quantum circuits with few trainable parameters and empirically testable operator roles. We evaluate three quantum generator architectures, referred to as ZZ, directed ZZ–CRX, and structured time evolution, within a hybrid Wasserstein quantum generative adversarial network trained to generate 16-step Bitcoin return windows. Across circuits with one to four layers and one to eight trainable coefficients, these generators recover meaningful marginal and temporal distributional features, while single-layer configurations remain competitive with deeper variants. Across matched depths, independently trained ZZ–CRX configurations exhibit a predominantly downward shift in median leverage relative to ZZ, consistent with the intended directional role of the controlled-rotation construction. The small number of coefficients also enables direct landscape analysis, revealing competing parameter preferences between marginal and temporal objectives and helping to explain the difficulty of finding parameter settings that perform well on both. These results provide an initial demonstration that structural assumptions can guide the design of compact quantum generators and link circuit architecture to observable statistical behaviour, motivating more systematic and parameter-efficient approaches to quantum circuit construction.
}


\maketitle

\section{Introduction}

Parameterised quantum circuits provide trainable models for variational
approaches to quantum computation \citep{cerezo2021variational}, with
applications to ground-state calculations and quantum simulation
\citep{peruzzo2014variational,li2017efficient}, combinatorial optimisation
\citep{farhi2014quantum}, and machine learning
\citep{mitarai2018quantum,schuld2020circuit}. Quantum generative approaches use
such circuits to represent and learn probability distributions, including
through quantum circuit Born machines and adversarial learning
\citep{liu2018differentiable,lloyd2018quantum}. 

In practice, circuit ansatzes are often selected for hardware convenience
or assembled from standard circuit templates without explicitly tailoring
their structure to the target data. These templates provide convenient
building blocks, but their use does not necessarily establish a connection
between the circuit operations and the mechanisms, symmetries, or
statistical properties relevant to the generation task. Increasing circuit
complexity can expand expressivity without guaranteeing improved task
performance, while making the roles of individual parameters harder to
interpret. Greater expressivity can also introduce optimisation
difficulties, including barren plateaus under certain circuit and
cost-function conditions, where gradients become difficult to
resolve
\citep{mcclean2018barren,cerezo2021cost,holmes2022connecting}.

Structured ansatz design provides a compelling alternative by imposing
a controlled restriction on the search space, favouring circuit terms
with explicit, testable roles. By connecting operator choices and
parameter-sharing rules to properties of the target data, this approach
offers a promising route to compact and interpretable quantum generators.
This design philosophy connects to a broader principle shared by classical and quantum machine learning: incorporating problem structure through architectural inductive biases. Examples include classical graph networks, equivariant quantum graph circuits, and geometric quantum machine learning, where relationships or symmetries guide model construction~\citep{mernyei2022equivariant,ragone2023representation,battaglia2018relational}.
These developments motivate asking whether the same design perspective can be
extended to quantum generation, with modelling assumptions informing circuit
architecture.

This work investigates a problem-informed approach to quantum generation by
constructing compact circuits from operators with explicit intended roles and
coefficients shared according to modelling assumptions. Here, interpretability
refers to explicit operator-role hypotheses, their comparison with observed
statistical responses, and direct inspection of low-dimensional parameter
landscapes.

We study this question using a quantum generative adversarial network (QGAN)
trained to generate financial return windows. Financial time series provide a
particularly demanding testbed because their marginal distributions are often
heavy-tailed while their temporal structure exhibits little linear return
autocorrelation, persistent volatility dependence, and asymmetric
leverage-like effects \citep{cont2001stylized,zhang2018cryptocurrency,ardia2019regime}. A successful generator must
therefore reproduce several distinct statistical targets rather than only a
single marginal distribution. This setting is well suited to testing whether
specific circuit terms and coefficient-sharing schemes contribute to identifiable
aspects of the generated dynamics.

The quantum generators in this work are designed according to three principles.
First, \emph{problem-informed structure}: circuit terms represent explicit
hypotheses about latent conditions, local interactions, and asymmetric temporal
modulation. Second, \emph{symmetry-inspired coefficient sharing}: coefficients
shared across interaction terms encode approximate homogeneity across
short-range interactions as an architectural inductive bias. Third, efficient
shallow construction: we prioritise generators with one or a few ansatz layers
and favour richer operator sets over additional layer repetitions to increase
expressivity.

These principles lead to three quantum generators for complete 16-step return windows: a ZZ-based construction, a directed ZZ–CRX extension, and a structured time-evolution construction. We implement these generators within the expectation-value Wasserstein QGAN framework of Dechant et al. (2026), extending it through problem-informed circuit design and a four-channel critic representation that exposes marginal and temporal features.

Our contributions are fourfold. First, we formulate a problem-informed design protocol that connects modelling assumptions to circuit operators and parameter-sharing rules. Second, we instantiate this protocol in three eight-qubit generators with one or two shared coefficients per layer and report their implementation-level resources. Third, we evaluate their marginal and temporal fidelity against hardware-efficient quantum and classical reference models. Finally, we examine training records and fixed-latent objective landscapes to assess the effects of layer repetition and characterise trade-offs between statistical targets.

\section{Problem formulation and data}
\label{sec:problem-formulation}

\subsection{Financial generation task and target statistics}

Let $r_t=\log P_t-\log P_{t-1}$ denote the daily Bitcoin log return. Following the financial time-series generation approach of \citet{dechant2025quantum}, we model the distribution of complete return windows rather than perform one-step forecasting. Here, we consider 16-step windows $\mathbf r_i=(r_{i,1},\ldots,r_{i,16})\in\mathbb R^{16}$. The model should reproduce both the distribution of individual return values and the local temporal relationships within each window.

The marginal target is motivated by the familiar shape of financial-return
distributions, where a concentrated centre coexists with extended tails
\citep{cont2001stylized}. The temporal targets describe three complementary
features. Raw returns should preserve the weak short-lag linear autocorrelation
associated with limited linear predictability. Absolute returns should retain
the positive short-range dependence associated with volatility clustering,
which is also widely reported for cryptocurrencies
\citep{zhang2018cryptocurrency}. Finally, the relation between a return and
future squared returns is used to examine leverage-like asymmetry. This last
signal is weaker and more sample dependent for Bitcoin, where inverse leverage
has also been reported \citep{ardia2019regime}. Accordingly, the analysis
evaluates leverage sign and magnitude as sample-specific properties of the
present Bitcoin series.

The length-16 output places a clear limit on what these temporal targets mean.
Our analysis measures the distribution of local within-window dependence
estimates over lags 1--4; it does not recover the autocorrelation structure of
an indefinitely generated or globally contiguous time series. Each estimate is
formed from only a small number of lagged pairs and consequently shows
substantial sampling variability. That variability is nevertheless part of the
data representation presented to the generator. A satisfactory model should
therefore reproduce not only the centre of each temporal statistic across
generated windows, but also its broad window-to-window distribution.

These requirements make the generation task multi-objective. We therefore
evaluate marginal shape, raw-return autocorrelation, absolute-return
autocorrelation, and leverage-like dependence separately, using the estimators
and comparison metrics defined in Section~\ref{sec:experimental-design}.

\subsection{Dataset, windowing, and preprocessing}

The empirical series contains 2,312 finite daily Bitcoin log returns from
2 January 2020 through 1 May 2026. Consecutive length-16 windows are extracted
with stride one, giving 2,297 overlapping windows. The overlap allows the model
to use all available local trajectories, but it also means that the windows
should not be interpreted as independent market episodes. All models are
trained and evaluated against this same historical series, with no held-out
market period. A \textit{fresh} evaluation uses newly sampled latent vectors
passed through the frozen generator.

We follow the preprocessing and inverse-mapping approach of \citet{dechant2025quantum}, with the transformations fitted to the Bitcoin series.
The procedure first
standardises the raw returns, applies the inverse Lambert-$W$ transform with tail
parameter $\delta=0.5$, standardises the transformed values again, and divides
them by their maximum absolute value. For inversion, the transformed-data
quantiles $0.0005$ and $0.9995$ are stored as bounds; generated values outside
this range are clipped before the remaining transformations are reversed.
Writing $\mathcal T$ for the fitted map and
$\widetilde{\bm r}=\mathcal T(\bm r)$, empirical and generated windows are
compared in the same bounded model space during training and returned to
financial units through $\mathcal T^{-1}$ for evaluation.

Because clipping acts directly on extreme generated values, the reported tail
behaviour describes the complete generation-and-inversion pipeline rather than
an unconstrained output distribution. Further implementation details are
provided in Online Resource~1, Section~S1.1.

\section{Wasserstein QGAN architecture}
\label{sec:common-wqgan}

\subsection{Hybrid adversarial formulation}

Generative adversarial networks learn an implicit data distribution through
competition between a generator and a discriminator
\citep{goodfellow2014generative}. The generator maps random latent variables to
synthetic samples, while the discriminator learns to separate those samples
from the training data. Quantum generative adversarial networks extend this
framework by implementing the generator, discriminator, or both with
parameterised quantum circuits
\citep{lloyd2018quantum,dallaire2018quantum,zoufal2019quantum}.

Following the hybrid architecture of \citet{dechant2025quantum}, we combine an expectation-value quantum generator with a classical convolutional critic.

The quantum circuit acts as an expectation-value sampler. Classical latent
variables are uploaded to the circuit, and Pauli expectation values form a
continuous generated window. This differs from a quantum circuit Born machine,
where individual samples are discrete measurement bitstrings drawn according to
the final-state probabilities. The expectation-value construction is therefore
well matched to continuous returns while retaining the latent variation needed
to generate different windows.

We train the model using the Wasserstein GAN formulation
\citep{arjovsky2017wasserstein}. By Kantorovich--Rubinstein duality, the
Wasserstein-1 distance between real and generated distributions can be written
as
\begin{equation}
 W_1(P_r,P_g)
 =
 \sup_{\lVert f\rVert_L\leq 1}
 \left[
 \mathbb E_{\bm u\sim P_r}f(\bm u)
 -
 \mathbb E_{\widetilde{\bm u}\sim P_g}f(\widetilde{\bm u})
 \right],
 \label{eq:wasserstein-dual}
\end{equation}
where the supremum is taken over all 1-Lipschitz functions. A WGAN replaces the
probabilistic discriminator with a critic that parameterises this scalar
function. The critic assigns higher scores to empirical windows than to
generated windows, while the generator changes its parameters to reduce the
score difference. A gradient penalty approximately enforces the Lipschitz
condition and supports more stable adversarial training
\citep{gulrajani2017improved}.

\subsection{Quantum generator and readout interface}

A latent vector $\bm z\sim\mathcal U[-\pi,\pi]^{16}$ is encoded into an
eight-qubit circuit. For trainable parameters $\bm\theta$, the generator is
read out through single-qubit Pauli expectation values, which are assembled
into a continuous 16-component generated window
$\bm y_{\bm\theta}(\bm z)\in[-1,1]^{16}$.

During training, $\bm y_{\bm\theta}$ is compared directly with the transformed
empirical window $\widetilde{\bm r}=\mathcal T(\bm r)$. For financial
diagnostics it is mapped back to the return window
\begin{equation}
 G_{\bm\theta}^{(r)}(\bm z)
 =
 \mathcal T^{-1}\!\left(\bm y_{\bm\theta}(\bm z)\right)
 \in\mathbb R^{16}.
 \label{eq:return-space-generator}
\end{equation}


\subsection{Four-channel critic representation and training objective}

Before entering the critic, a model-space window $\bm u$ is represented through
four aligned channels,
\begin{equation}
 \Phi(\bm u)=
 \bigl[
 \bm u,\ |\bm u|,\ \bm u^{\odot 2},\ \bm v^{(5)}(\bm u)
 \bigr]
 \in\mathbb R^{16\times4},
 \label{eq:critic-channels}
\end{equation}
where $\bm u^{\odot2}$ denotes elementwise squaring and $\bm v^{(5)}$ is the
five-step trailing population standard deviation of the model-space returns,
including the current observation; see Online Resource~1, Section~S1.2.
Unavailable initial entries are padded
by repeating the first return, so no information crosses a window boundary.
The raw channel retains sign and
direction, the absolute and squared channels expose magnitude dependence, and
the rolling channel supplies local scale information. All quantum and classical
generators are evaluated through this same representation.

Writing $P_{\widetilde r}$ for the empirical model-space window distribution and
$P_z$ for the latent distribution, the critic approximately maximises
\begin{equation}
\begin{split}
 \mathcal L_D={}&
 \mathbb E_{\widetilde{\bm r}\sim P_{\widetilde r}}
       [D_{\bm\phi}(\Phi(\widetilde{\bm r}))]
 -\mathbb E_{\bm z\sim P_z}
       [D_{\bm\phi}(\Phi(\bm y_{\bm\theta}(\bm z)))]\\
 &-\lambda_{\mathrm{GP}}\,
  \mathbb E_{\widehat{\bm r}}
  \left(
    \left\lVert
      \nabla_{\widehat{\bm r}}
      D_{\bm\phi}(\Phi(\widehat{\bm r}))
    \right\rVert_2-1
  \right)^2 ,
\end{split}
\label{eq:wqgan-critic-objective}
\end{equation}
where $\widehat{\bm r}$ interpolates between an empirical and a generated
window in the 16-dimensional model space. The gradient is therefore taken with
respect to the interpolated window through the four-channel transformation. The
generator minimises
\begin{equation}
 \mathcal L_G
 =
 -\mathbb E_{\bm z\sim P_z}
 \left[
 D_{\bm\phi}(\Phi(\bm y_{\bm\theta}(\bm z)))
 \right].
 \label{eq:wqgan-generator-objective}
\end{equation}

The critic-score difference approaches the dual Wasserstein objective only to the extent that the finite critic class is sufficiently optimised under the Lipschitz constraint. We therefore use it as an adversarial training signal, while checkpoint selection and evaluation rely on the critic-independent diagnostics described in Section~\ref{sec:model-selection-evaluation}.

\section{Problem-informed quantum generator design}
\label{sec:problem-informed-design}
\subsection{Design viewpoint: encoding temporal structure in quantum circuits}

A financial time series partitioned into short observation windows is modelled
here as an ensemble of short trajectories with varying window-specific latent
conditions. Over a market regime that is long relative to the 16-step window,
we hypothesise that the dominant short-range interaction rules vary on a longer
timescale than individual return fluctuations and are approximately homogeneous
across nearby temporal positions. These assumptions guide the circuit
construction. Each window is represented through ordered local readouts.
Financial time labels these output positions, whereas circuit evolution
constructs their joint dependence. Nearest-neighbour couplings encode the
assumed local interaction structure, while shared coefficients express its
approximate temporal homogeneity.

This approach follows the classical principle of relational inductive biases:
architectural choices encode assumptions about entities, their interactions,
and shared rules \citep{battaglia2018relational}. Related quantum approaches
include equivariant quantum graph circuits and geometric quantum machine
learning, which incorporate relational structure and symmetry into circuit
design \citep{mernyei2022equivariant,ragone2023representation}. Here,
parameter sharing encodes approximate temporal homogeneity as an architectural
prior.

The operator choices additionally reflect hypotheses about the form of
temporal dependence. Pair couplings provide local interactions, while directed
controlled rotations introduce state-dependent asymmetry motivated by
leverage-like dependence. These choices define a phenomenological model whose
intended statistical effects are tested through marginal and temporal
diagnostics.

\begin{figure}[!tbp]
  \centering
  \includegraphics[width=\textwidth]{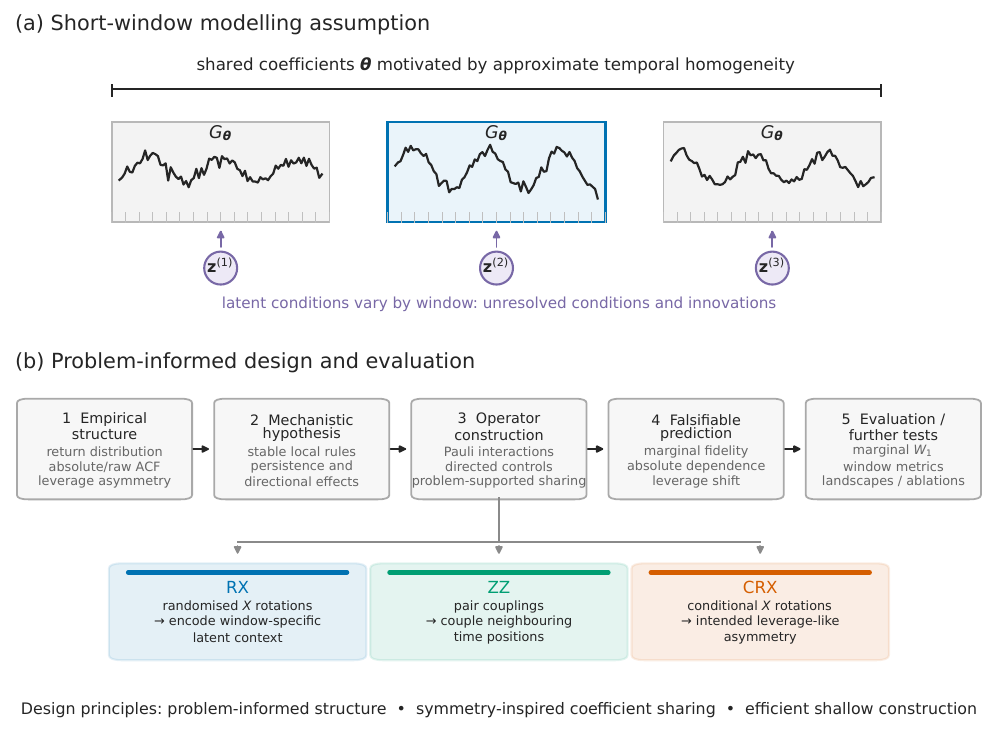}
  \caption{Problem-informed generator design. \textbf{(a)} Window-specific
  latent variables $\mathbf{z}^{(w)}$ represent unresolved conditions and
  innovations, while shared coefficients $\boldsymbol{\theta}$ encode the
  approximate temporal-homogeneity prior. \textbf{(b)} The design and testing
  workflow maps empirical structure to operator-role hypotheses and evaluates
  their statistical consequences. Ablations are shown as further tests.}
  \label{fig:design-protocol}
\end{figure}

\subsection{Single-observable design for the Z readout}

We first formulate the design for the single-observable sequence
$\mathbf y_Z=(\langle Z_0\rangle,\ldots,\langle Z_7\rangle)$, whose consecutive
positions correspond to consecutive qubits. The construction combines latent
encoding, local pair interactions, and directed conditional modulation.
Section~\ref{sec:dual-observable-extension} extends these blocks to the grouped
$X$ and $Z$ readouts.

\subsubsection{RX latent encoding}

For an eight-dimensional latent vector $\mathbf{z}^{Z}$, the idealised
single-readout encoding is

\begin{equation}
U_{\mathrm{enc}}^{Z}(\mathbf{z}^{Z})
=
\prod_{i=0}^{7}R_X^{(i)}(z_i^{Z}).
\label{eq:z-encoding}
\end{equation}

Acting on the initialised qubits, the $R_X$ rotations encode latent variation
in the computational-basis populations and hence in the $Z$ expectations. The
latent angles are sampled to represent window-specific conditions and
innovations; they are not inferred from observed pre-window histories.

\subsubsection{ZZ pair coupling}

We introduce nearest-neighbour $ZZ$ rotations as local interaction terms in the
single-observable construction,

Throughout, $R_{AB}^{(i,j)}(\alpha)=\exp[-i\alpha A_iB_j/2]$, with
$A,B\in\{X,Y,Z\}$, where $i$ and $j$ identify the two acted-on qubits and
identity operators on all other qubits are implicit.

\begin{equation}
U_{\mathrm{pair}}^{Z}(\alpha_\ell)
=
\prod_{i=0}^{6}R_{ZZ}^{(i,i+1)}(\alpha_\ell).
\label{eq:z-pair-unitary}
\end{equation}

The coefficient $\alpha_\ell$ is shared across neighbouring pairs within layer
$\ell$. An isolated $ZZ$ rotation commutes with the local $Z$ observables and
therefore leaves their expectations unchanged; its influence on the generated
$Z$-readout sequence arises through composition with noncommuting operations.

\subsubsection{CRX directed control}

To motivate a directed interaction, consider the single idealised term

\begin{equation}
H_{i\rightarrow i+1}
=
H_x(I_i+Z_i)X_{i+1}
=
2H_xP_i^{(0)}X_{i+1},
\qquad
P_i^{(0)}=\frac{I_i+Z_i}{2}.
\label{eq:ideal-crx-hamiltonian}
\end{equation}

This term motivates $C_0R_X$, an $R_X$ target rotation controlled on
$\lvert0\rangle$, rather than the standard $\lvert1\rangle$-controlled $R_X$
gate. When the control is $\lvert0\rangle_i$, the conditional field is active
and the target evolves under $2H_xX_{i+1}$; when the control is
$\lvert1\rangle_i$, this single term vanishes. We set $\hbar=1$. Under the
implemented convention $R_X(\beta)=\exp(-i\beta X/2)$, the active-branch angle
is $\beta=4H_x\tau$. For a target initially aligned with the $Z$ axis, let
$\langle Z_{i+1}(0)\rangle=s$, where $s=+1$ for $\lvert0\rangle$ and $s=-1$
for $\lvert1\rangle$. In the active branch,

\begin{equation}
\langle Z_{i+1}(\tau)\rangle
=
s\cos(4H_x\tau).
\label{eq:crx-z-aligned}
\end{equation}

For sufficiently short evolution time, the active branch moves either
initially $Z$-aligned target toward zero $Z$ expectation. In the idealised
readout interpretation, positive control polarisation activates this
contraction, while the negative branch leaves the target unchanged under the
isolated term. This relative suppression motivates the controlled rotation as
a candidate mechanism for leverage-like asymmetry. Its persistence in the
complete generator is an empirical question; Online Resource~1,
Section~S2 gives the full evolution and the
conditions for contraction.

\subsection{Dual-observable extension and boundary terms}
\label{sec:dual-observable-extension}

The complete generator produces a 16-component window from the $X$ and $Z$
expectations of eight qubits, grouped as follows.

\begin{equation}
\mathbf{y}
=
\left(
\langle X_0\rangle,\ldots,\langle X_7\rangle,
\langle Z_0\rangle,\ldots,\langle Z_7\rangle
\right).
\label{eq:generator-output}
\end{equation}

Within each group, consecutive output positions correspond to consecutive
qubits. This ordering places the two readouts from the same qubit eight
positions apart.

The second readout branch is obtained through the basis-dual construction

\begin{equation}
R_X\longleftrightarrow R_Z,
\qquad
R_{ZZ}\longleftrightarrow R_{XX},
\qquad
CR_X\longleftrightarrow CR_Z.
\label{eq:basis-dual-map}
\end{equation}

Accordingly, for latent vector $\mathbf{z}\in\mathbb{R}^{16}$, the implemented
encoding is

\begin{equation}
U_{\mathrm{enc}}(\mathbf{z})
=
\prod_{i=0}^{7}
R_Z^{(i)}(z_{2i+1})R_X^{(i)}(z_{2i}),
\label{eq:latent-encoding}
\end{equation}

where the rightmost rotation acts first. The two rotations specify
complementary latent-encoding directions for the two readout bases.

The implemented pair block is

\begin{equation}
U_{\mathrm{pair}}(\alpha_\ell)
=
\left[
\prod_{i=0}^{6}R_{ZZ}^{(i,i+1)}(\alpha_\ell)
\right]
R_{XZ}^{(7,0)}(\alpha_\ell)
\left[
\prod_{i=0}^{6}R_{XX}^{(i,i+1)}(\alpha_\ell)
\right].
\label{eq:pair-unitary}
\end{equation}

The $R_{XX}$ and $R_{ZZ}$ terms form the basis-dual pair construction. The
mixed $R_{XZ}^{(7,0)}$ term is assigned to the concatenation boundary between
$\langle X_7\rangle$ and $\langle Z_0\rangle$. It belongs to this output
ordering rather than to a homogeneous ring closure. As with the $ZZ$ terms,
its effect on these readouts depends on its composition with noncommuting
operations. In each nearest-neighbour sweep, the implementation applies pairs
in ascending index order, from $(0,1)$ through $(6,7)$.

Define also $P_i^{(+)}=(I_i+X_i)/2$. The operator content of the implemented
directed block is

\begin{equation}
G_{\mathrm{dir}}
=
\frac{1}{2}
\left[
\sum_{i=0}^{6}P_i^{(+)}Z_{i+1}
+P_7^{(+)}X_0
+\sum_{i=0}^{6}P_i^{(0)}X_{i+1}
\right].
\label{eq:directed-generator}
\end{equation}

Equation~\eqref{eq:directed-generator} summarises the constituent generators;
Eq.~\eqref{eq:directed-unitary} specifies their ordered implementation. Since
the constituent terms generally do not commute, the product should not be
identified with the exact exponential of their sum. Here $C_{0}R_X$ is an
$R_X$ target rotation controlled on $\lvert0\rangle$, $C_{+}R_Z$ is an $R_Z$
target rotation controlled on $\lvert+\rangle$, and the boundary operation
$C_{+}R_X$ is an $R_X$ target rotation controlled on $\lvert+\rangle$. The
implemented order is

\begin{equation}
U_{\mathrm{dir}}(\beta_\ell)
=
\left[\prod_{i=0}^{6}C_{0}R_X^{(i\rightarrow i+1)}(\beta_\ell)\right]
C_{+}R_X^{(7\rightarrow0)}(\beta_\ell)
\left[\prod_{i=0}^{6}C_{+}R_Z^{(i\rightarrow i+1)}(\beta_\ell)\right].
\label{eq:directed-unitary}
\end{equation}

The rightmost controlled-$R_Z$ sweep is applied first, followed by the boundary
operation and then the controlled-$R_X$ sweep. Within each sweep, controls are
applied in ascending index order from $0\rightarrow1$ through
$6\rightarrow7$, so later operations appear farther to the left in the
unitary product. We retain CRX in the architecture name because the controlled
X rotation motivates the Z-readout mechanism; the dual-observable
implementation also includes its controlled Z counterpart.

\begin{figure}[!tbp]
  \centering
  \includegraphics[width=\textwidth]{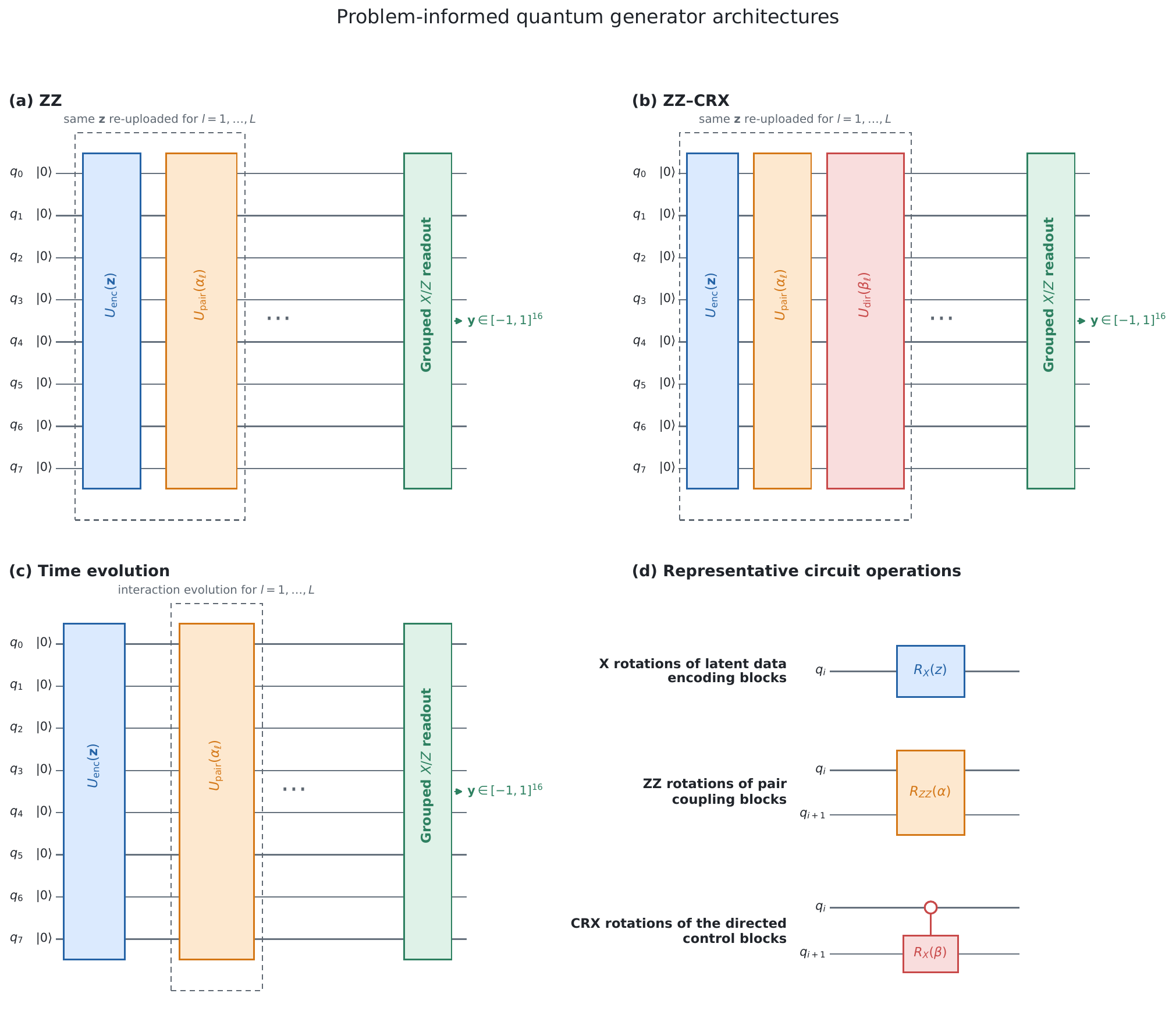}
  \caption{Problem-informed quantum generator architectures. All three
  generators act on eight qubits and produce the grouped expectation-value
  vector in Eq.~\eqref{eq:generator-output}. Dashed enclosures show the layer
  schedules for \textbf{(a)} ZZ, \textbf{(b)} ZZ--CRX, and \textbf{(c)} Time
  evolution. \textbf{(d)} Representative circuit operations; basis-dual and
  mixed-basis boundary operations are omitted for visual economy.}
  \label{fig:circuit-architectures}
\end{figure}

\subsection{Assembly of the generator architectures}
\label{sec:architecture-assembly}

The three proposed architectures combine the complete blocks using different
schedules. In each left-arrow product below, layer 1 acts first and larger
layer indices therefore appear farther to the left. The ZZ generator
re-uploads the same latent vector before each pair-coupling block,

\begin{equation}
U_{\mathrm{ZZ}}^{(L)}
=\prod_{\ell=1}^{L}
U_{\mathrm{pair}}(\alpha_\ell)U_{\mathrm{enc}}(\mathbf{z}).
\label{eq:zz-architecture}
\end{equation}

The ZZ--CRX generator extends the pair-only architecture to examine the
directional hypothesis for leverage-like dependence. It inserts the
directed-control block in each layer,

\begin{equation}
U_{\mathrm{ZZ\text{-}CRX}}^{(L)}
=
\prod_{\ell=1}^{L}
U_{\mathrm{dir}}(\beta_\ell)
U_{\mathrm{pair}}(\alpha_\ell)
U_{\mathrm{enc}}(\mathbf{z}).
\label{eq:zz-crx-architecture}
\end{equation}

\noindent\begin{minipage}{\textwidth}
The time-evolution family uses one latent encoding followed by repeated
structured pair-interaction blocks,

\begin{equation}
U_{\mathrm{TE}}^{(L)}
=
\left[
\prod_{\ell=1}^{L}
U_{\mathrm{pair}}(\alpha_\ell)
\right]U_{\mathrm{enc}}(\mathbf{z}).
\label{eq:time-evolution-architecture}
\end{equation}

Here, time evolution denotes the ordered sequence of pair-interaction blocks
in Eq.~\eqref{eq:time-evolution-architecture}, with each layer coefficient
trained independently.
\end{minipage}

The one-layer ZZ and time-evolution circuits coincide by construction. The
trainable parameters are the layer coefficients $\{\alpha_\ell\}$ for ZZ and
time evolution, and $\{\alpha_\ell,\beta_\ell\}$ for ZZ--CRX. Their parameter
budgets are compared in Section~\ref{sec:parameter-budgets}; full architectural
details and implementation-level resource counts are provided in Online Resource~1,
Table~S4.

The resulting operator-role hypotheses guide the marginal and temporal
comparisons below.

\section{Experimental design and evaluation}
\label{sec:experimental-design}

\subsection{Reference models and comparison rationale}

The experiments are organised around the three problem-informed generators introduced in Section 4. All generators are trained against the same four-channel critic and Wasserstein objective, so the comparison focuses on the impact of generator architecture. All quantum generators are trained and evaluated using noiseless state-vector simulation. Pauli expectation values are computed directly from the simulated states, without finite-shot measurement sampling or a hardware-noise model. The common optimisation schedule and numerical settings are provided in Online Resource~1, Section~S1.

The three problem-informed generators and a hardware-efficient ansatz (HEA) are evaluated with eight qubits at depths 1--4. The HEA serves as a domain-agnostic quantum benchmark, providing variational flexibility without explicitly encoding hypotheses about the target financial statistics. A two-parameter classical generator applies a shared nonlinear map combining each latent variable with its nearest-neighbour mean under circular boundaries. This control tests whether a similarly compact classical construction with parameter sharing can reproduce the observed performance. Temporal convolutional neural network (TCNN) generators provide more flexible classical references, with hidden widths of 2, 4, 6, and 8 corresponding to 29, 81, 157, and 257 active parameters. These models assess marginal and temporal fidelity across a broader classical parameter budget. Together, the benchmarks place the proposed architectures in context with respect to circuit structure, parameter sharing, and model capacity; they do not define an exhaustive or fully optimised classical frontier.

\subsection{Model selection and statistical evaluation}
\label{sec:model-selection-evaluation}

Model selection uses marginal criteria exclusively. For each quantum architecture--depth configuration, three runs are trained with different seeds. Marginal $W_1$ is monitored during training using 1,000 generated windows, and the three runs are ranked by their saved best-checkpoint monitoring values; the middle-ranked run is retained. For each problem-informed family, the representative depth is the one with the smallest marginal $W_1$ in the reported evaluation. All reported comparisons use 2,297 windows. Temporal diagnostics do not enter checkpoint, run, or configuration selection.

Marginal fidelity is assessed using the one-dimensional $W_1$ distance between the original 2,312 empirical daily log returns and the 36,752 pooled generated return values. Q–Q curves and their root-mean-square deviation from the equality line provide complementary diagnostics. Temporal diagnostics instead preserve complete 16-step windows and evaluate raw-return autocorrelation, absolute-return autocorrelation, and leverage-like dependence at lags 1--4. These statistics use globally centred, finite-window-corrected normalised covariances. Lag curves report the median across window-level estimates, and scalar temporal errors average the absolute empirical--generated median difference over the four lags. No lagged pair crosses a window boundary. Full estimator definitions are provided in Online Resource~1, Section~S1.5.

Uncertainty in the temporal median curves is summarised by 95\% simultaneous bootstrap intervals over lags 1--4. For the empirical data, contiguous returns are resampled in blocks before reconstructing overlapping windows; generated data are resampled as complete windows. These procedures account for the different dependence structures of the two sources. Bootstrap settings and additional sampling-resolution analyses are provided in Online Resource~1, Section~S1.6.

\subsection{Loss-landscape analysis}
\label{sec:loss-landscape-analysis}

To examine whether marginal fidelity and short-window absolute-return dependence favour the same parameter regions, we scan three configurations with two active coefficients: ZZ at depth 2, ZZ--CRX at depth 1, and time evolution at depth 2. At each parameter pair, we evaluate marginal $W_1$ and the mean over lags 1--4 of the Wasserstein distances between empirical and generated window-level absolute-return-ACF distributions. This temporal objective compares the distributions of window-level estimates, complementing the median-based errors used in the cross-model evaluation.

We also examine the temporal landscapes separately at lags 1--4 to assess whether averaging conceals competing parameter preferences across lags (Online Resource~1, Figure~S7). These surfaces use the same parameter grids and fixed latent samples as the lag-averaged analysis. A fixed latent batch is reused across each grid to reduce Monte Carlo variation between parameter settings. Stored checkpoints are overlaid to locate the trained configurations on these diagnostic surfaces. Grid settings, checkpoint mapping, and the detailed per-lag analysis are provided in Online Resource~1, Sections~S5--S5.3.

The displayed domains follow the output periods of the active coefficients. For
ZZ and time evolution, $(\alpha_1,\alpha_2)$ each span
$[-\pi/2,3\pi/2)$ on a $50\times50$ grid. For ZZ--CRX, $\alpha_1$
spans $[-\pi/2,3\pi/2)$ while the directed coefficient $\beta_1$ spans
$[-\pi,3\pi)$ on a $50\times100$ grid. Thus all axes use the common
angular step $\pi/25$, while the ZZ--CRX vertical axis covers its full
$4\pi$ period rather than a relabelled $2\pi$ interval.

\section{Results}

\subsection{Marginal distribution fidelity}

\begin{figure}[!tbp]
  \centering
  \includegraphics[width=\textwidth]{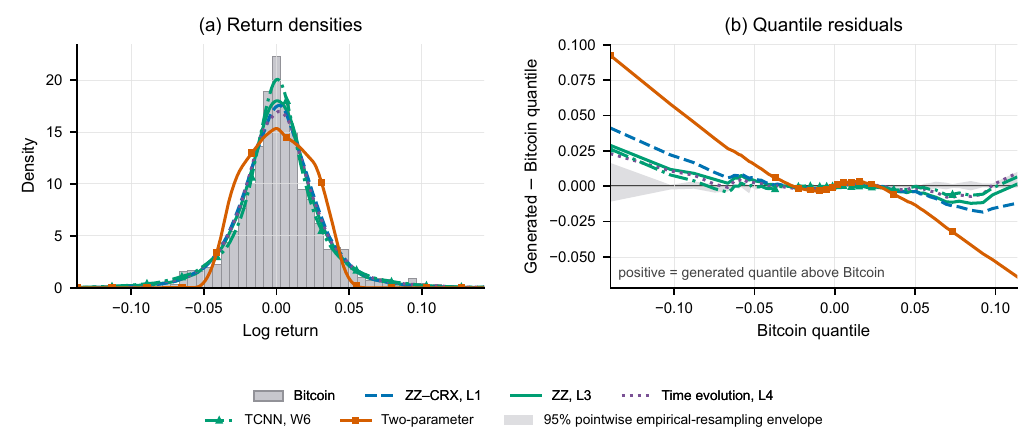}
  \caption{Representative marginal diagnostics. \textbf{(a)}
  Return densities for Bitcoin and the selected representatives ZZ L3/run 2,
  ZZ--CRX L1/run 1, time-evolution L4/run 0, TCNN W6/run 2, and the
  two-parameter classical control/run 1. \textbf{(b)} Generated minus Bitcoin
  quantiles on the common 240-point probability grid. The grey region is a
  95\% pointwise empirical-resampling envelope from 10,000 fixed-seed
  complete-window resamples of 2,297 windows each. It describes variability
  under window resampling, without accounting for serial dependence between
  overlapping historical windows. This figure reports representative
  configurations under the marginal-only selection rule and is not a formal
  cross-architecture goodness-of-fit test.}
  \label{fig:marginal-representatives}
\end{figure}

Figure~\ref{fig:marginal-representatives} summarises the marginal comparison for each selected representative. In panel~(a), the three problem-informed generators recover the concentrated central mass and span much of the observed return range, although their peak heights and tail densities do not exactly match the empirical distribution.
Among them, the time-evolution layer-4 representative gives the closest marginal agreement, whereas the two-parameter classical control produces an overly compressed distribution and misses substantial portions of both tails. TCNN W6 provides the closest displayed density match.
The quantile residuals in Figure~\ref{fig:marginal-representatives}(b) make the remaining discrepancies more visible. The problem-informed representatives remain relatively close to zero through much of the central quantile range, with larger departures toward the tails. Their marginal \(W_1\) values range from 0.00176 to 0.00249, and their Q--Q RMSE values range from 0.00256 to 0.00466 (Table~\ref{tab:representative-performance}). Time evolution gives the clearest result, reaching \(W_1=0.00176\) and a Q--Q RMSE of 0.00256. TCNN W6 performs better numerically, with \(W_1=9.3\times10^{-4}\), although its residual curve still leaves the empirical-resampling envelope over parts of the tails.

Online Resource~1, Figures~S2 and S3, extend this comparison across all circuit depths and TCNN widths. These figures show that the broad marginal agreement is not uniform: fidelity varies across architectures and depths, and some configurations exhibit appreciable shape distortion, including the excessive central concentration of ZZ at layer 2. Together with the corresponding across-seed training histories, these results indicate that marginal fidelity is sensitive to both circuit depth and optimisation outcome. The selected representatives nevertheless reproduce the broad empirical shape rather than merely its location or scale.
Because these diagnostics pool returns across windows, they assess one-dimensional marginal fidelity but not the ordering or dependence of returns within each generated window.

\subsection{Temporal structure}

\begin{figure}[!tbp]
  \centering
  \includegraphics[width=\textwidth]{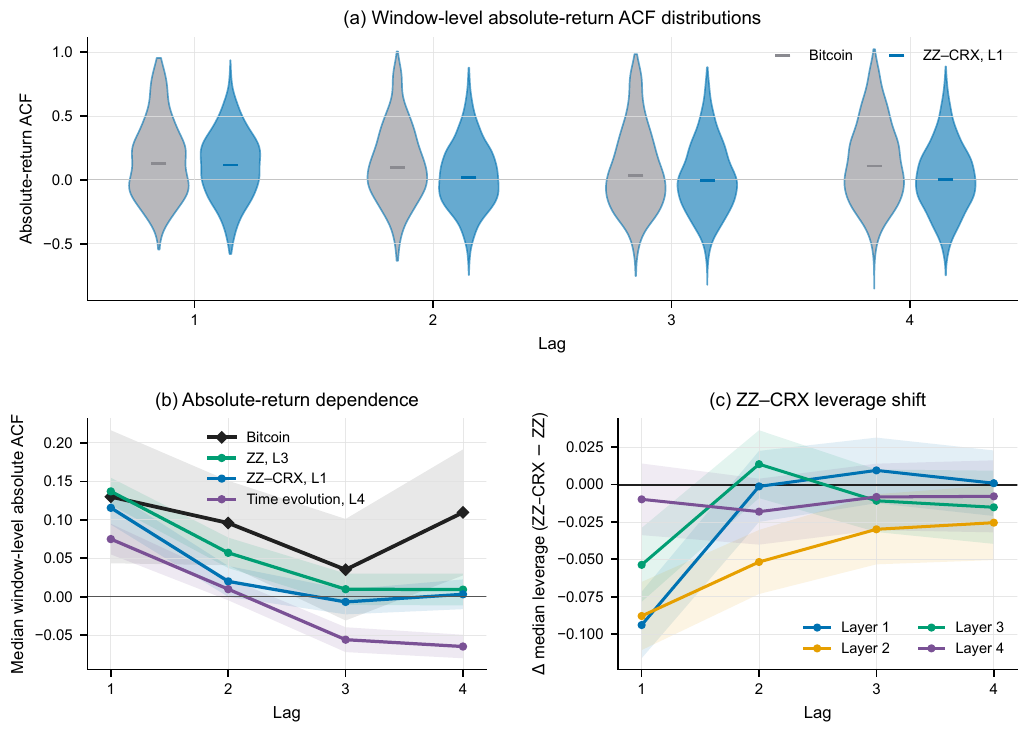}
  \caption{Short-window temporal structure. \textbf{(a)} Window-level
  absolute-return ACF distributions for Bitcoin and ZZ--CRX L1 at lags 1--4;
  violins show densities and horizontal segments mark medians. \textbf{(b)}
  Median absolute-return ACF for Bitcoin and the selected ZZ L3, ZZ--CRX L1,
  and time-evolution L4 representatives. \textbf{(c)} Differences in median
  leverage between ZZ--CRX and ZZ at matched depths; negative values indicate
  lower leverage under ZZ--CRX. Shading shows 95\% simultaneous max-$t$
  bootstrap intervals over the four lags, for the medians in \textbf{(b)} and
  the differences in \textbf{(c)}. Configurations were selected using marginal
  criteria. The leverage comparison uses independently trained runs and is not
  a controlled ablation.}
  \label{fig:temporal-mechanisms}
\end{figure}

Figure~\ref{fig:temporal-mechanisms} presents the temporal diagnostics at two
complementary levels. Panel (a) shows the window-level absolute-return ACF
distributions for Bitcoin and the selected ZZ--CRX layer-1 generator, while
panel (b) compares their medians with the other problem-informed
representatives. Online Resource~1, Figure~S4,
extends this comparison across depths and benchmark models. The broad violins
show that each median summarises considerable variation among windows; as
defined in Section~\ref{sec:problem-formulation}, these diagnostics describe
short-horizon, within-window structure.

Among the three temporal properties considered, absolute-return dependence
provides the clearest positive signal. The selected ZZ and ZZ--CRX
representatives capture the lag-1 value relatively closely, but underestimate
dependence at subsequent lags. Both show a weak upturn at lag 4, resembling the
direction of the empirical rebound without recovering its magnitude, whereas
the time-evolution representative departs more strongly at later lags.
Online Resource~1, Figure~S4, places this
agreement in context: ZZ and ZZ--CRX reproduce the initial positive dependence
more consistently than several benchmarks, particularly TCNN, whose
absolute-return ACF is generally weaker. The comparison remains
depth-dependent, however, and some HEA configurations also capture this
behaviour well.

Panel (c) compares median leverage between independently trained and selected
ZZ--CRX and ZZ configurations at matched depths. ZZ--CRX has lower median
leverage in 13 of the 16 matched depth--lag comparisons, with a mean
ZZ--CRX-minus-ZZ difference of $-0.02435$. Under this sign convention, negative
differences mean that ZZ--CRX has a lower median than ZZ at the same depth and
lag. This predominantly downward shift is consistent with the intended
directional role of the combined controlled-\(R_X\)/controlled-\(R_Z\)
construction. Its direction alone does not establish closer agreement with
empirical leverage. Online Resource~1,
Figure~S11, offers a
complementary view along training: ZZ L1's leverage discrepancies remain nearly
stationary once marginal fitting has saturated, whereas the CRX-containing runs
continue to explore different leverage behaviour, including during intervals of
relatively stable marginal fidelity. The two-layer ZZ--CRX run also reaches
lower lag-1 leverage discrepancy than the one-layer run. These observations
suggest additional temporal flexibility along training, although that
flexibility does not guarantee better agreement at every lag. Temporal
discrepancies can also continue to change beyond the minimum-recorded
marginal-\(W_1\) checkpoint, with improvements at some stages or lags
accompanied by deterioration at others. The selected checkpoints therefore
need not be optimal for temporal fidelity.

The empirical leverage uncertainty bands include zero at all four lags, and
classical models---especially TCNN W4---also closely follow the empirical
median curve (Online Resource~1, Figure~S6).
Together with the raw-return ACF comparisons in Online Resource~1,
Figure~S5, this illustrates why no single
diagnostic gives a complete picture of temporal fidelity.

\subsection{Marginal and temporal loss landscapes}

\begin{figure}[!tbp]
  \centering
  \includegraphics[width=\textwidth]{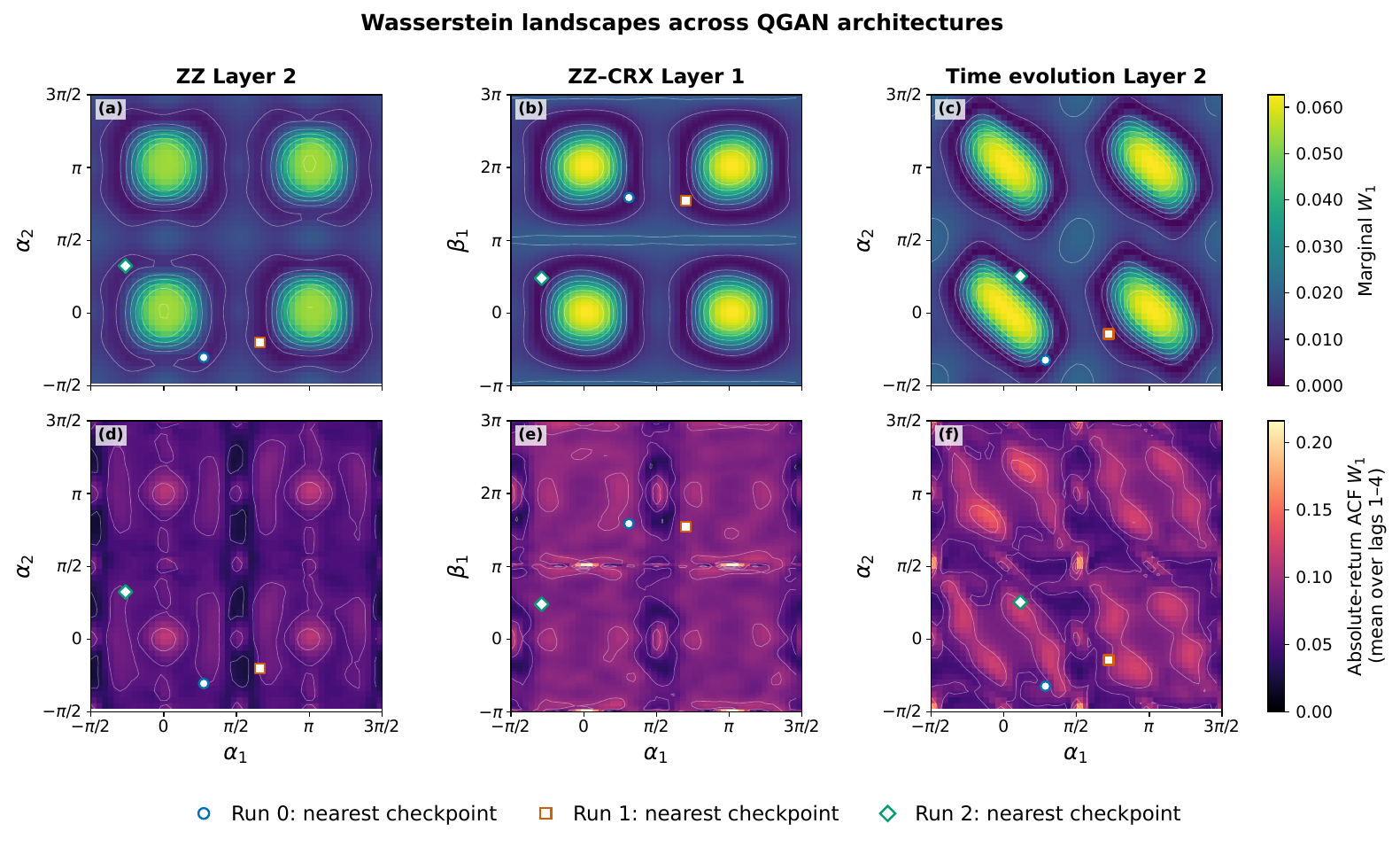}
  \caption{Marginal and temporal-objective landscapes for ZZ layer 2,
  ZZ--CRX layer 1, and time evolution layer 2. The upper row shows marginal
  $W_1$ for flattened returns; the lower row shows the mean over lags 1--4
  of the per-lag $W_1$ distances between empirical and generated
  window-level absolute-return-ACF distributions. For ZZ and time evolution,
  the coordinates $(\alpha_1,\alpha_2)$ each span $[-\pi/2,3\pi/2)$ on a
  $50\times50$ grid. For ZZ--CRX, $\alpha_1$ spans
  $[-\pi/2,3\pi/2)$ and $\beta_1$ spans $[-\pi,3\pi)$ on a
  $50\times100$ grid, covering its full $4\pi$ directed-angle period at the
  same angular resolution. Each configuration uses a fixed latent batch of
  1,000 windows throughout its grid. Checkpoint coordinates are mapped into
  the displayed domains using the respective $2\pi$ and $4\pi$ output
  periods. The circle, square, and diamond mark the saved
  checkpoints nearest the lowest recorded marginal-$W_1$ epoch of runs
  0, 1, and 2, respectively. Markers are overlays; colour scales are
  shared within each row.}
  \label{fig:wasserstein-landscapes}
\end{figure}

Figure~\ref{fig:wasserstein-landscapes} shows the marginal and lag-averaged
temporal objectives over the two parameters of three selected problem-informed
configurations.

Reading the upper and lower rows together reveals limited alignment
between marginal and temporal objectives within each configuration.
The ZZ--CRX panels retain the same physical width and height as the other
columns but cover twice the vertical angular range, so geometric distances are
interpreted from the coefficient ticks rather than from panel-space distances.
The low-value regions and sampled minima occupy visibly different parameter
regions. A move that improves marginal $W_1$ may
therefore leave absolute-return-ACF agreement unchanged or make it worse.
The four-channel Wasserstein critic can respond to both return and
temporal features during training, but the differing landscapes make a
parameter choice favourable to both diagnostics difficult to find.

The lag-averaged lower row also compresses conflicts among lags.
Online Resource~1, Figure~S7, shows
favourable regions shifting, splitting, or becoming unfavourable from one
lag to another. For ZZ--CRX, the lag-1 and lag-averaged temporal basins occupy
a region distinct from the preferred regions at the other lags. No common
low-discrepancy region is visually apparent across the marginal and all four
lag-specific landscapes.

The three markers in each column identify the saved checkpoints nearest
each run's lowest recorded marginal-$W_1$ epoch. Their locations reinforce the
qualitative pattern: checkpoints selected for favourable marginal performance
do not consistently lie within the most favourable temporal regions. This is
consistent with the marginal-only selection rule rather than convergence to a
joint compromise. Moving into a temporal basin can raise marginal loss, so
selecting the checkpoint by marginal $W_1$ need not retain the best temporal
state.

The ZZ and ZZ--CRX landscapes display approximate sign-reversal and reflection
patterns linked to their circuit structure. The single-observable identities
explain parts of this geometry; in the implemented dual-observable circuits,
the corresponding reflection patterns are empirical approximations. The
diagonal features of time evolution are consistent with partial accumulation
of its interaction angles.
Online Resource~1, Section~S5.3, gives the supporting
derivations and the limits of these interpretations.

\subsection{Single-layer competitiveness}

\begin{figure}[!htbp]
  \centering
  \includegraphics[width=\textwidth]{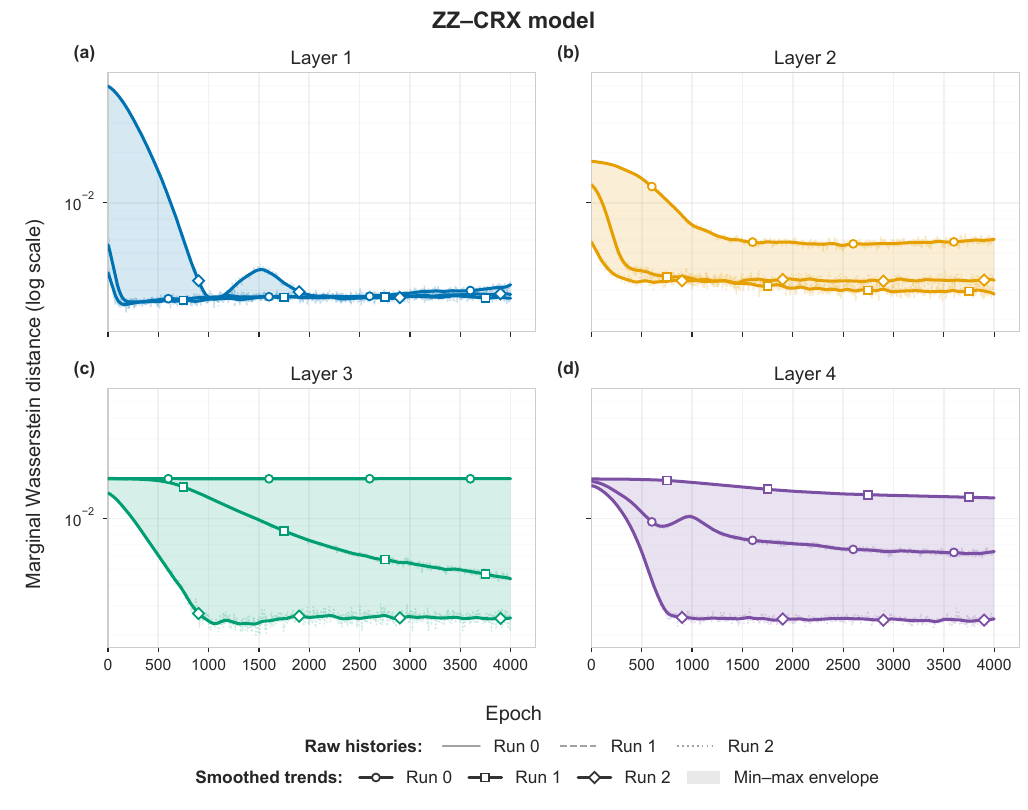}
  \caption{Marginal-Wasserstein training histories for the problem-informed
  ZZ--CRX model at \textbf{(a)} layer 1, \textbf{(b)} layer 2,
  \textbf{(c)} layer 3, and \textbf{(d)} layer 4. Every panel retains the
  unsmoothed histories of runs 0--2 as faint curves and shows their individual
  centred log-Gaussian trends. The shaded pointwise minimum--maximum envelope
  describes across-seed sensitivity; it is not a confidence interval or an
  across-run mean.}
  \label{fig:zz-crx-training}
\end{figure}

Figure~\ref{fig:zz-crx-training} follows the marginal-Wasserstein training
histories of the ZZ--CRX model from one to four layers. All three one-layer runs
reach a similar low-loss regime at parameter settings related approximately by
sign reversal and $\pi$ reflection, whereas added layers do not reliably lower
the recorded loss and show greater across-seed variation. Some three- and four-layer
runs remain at substantially higher loss. The selected ZZ--CRX representative
also uses one layer (Table~\ref{tab:representative-performance}); its reported
marginal $W_1$ is 0.00249, compared with
0.00612 for the four-layer representative
(Figure~\ref{fig:parameter-budget}). The recorded training curves and fresh
diagnostics are distinct measures, but both support the competitiveness of the
single-layer ZZ--CRX configuration.

Online Resource~1, Figures~S8 and S9, provide the corresponding
ZZ and time-evolution depth sweeps. Their one-layer configurations remain close
to the best-depth fresh marginal $W_1$ values within their respective families
(Table~S3). Thus,
the evidence supports deliberately shallow circuits as competitive starting
points for these problem-informed models; added depth improves particular
metrics or runs but produces no consistent gain across the three families.

\subsection{Performance under constrained parameter budgets}
\label{sec:parameter-budgets}

\begin{figure}[!tbp]
  \centering
  \includegraphics[width=\textwidth]{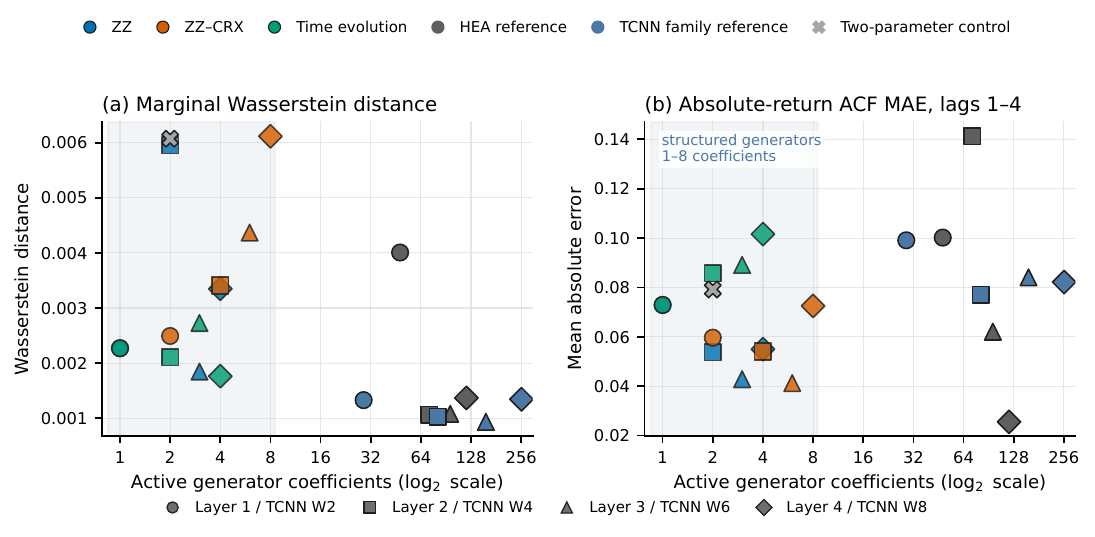}
  \caption{Performance versus active generator coefficient count for the
  problem-informed generators and reference benchmarks under the marginal and
  absolute-return-dependence objectives. \textbf{(a)} Marginal
  Wasserstein distance between empirical and generated returns. \textbf{(b)}
  Mean absolute error between empirical and generated median absolute-return
  ACF over lags 1--4. Lower values are better. Colour identifies generator
  family, while marker shape identifies quantum-circuit depth or TCNN width.
  Points show the selected representative run for each configuration
  (Section~\ref{sec:model-selection-evaluation}). Coefficient count measures
  parameter dimension, not total computational cost.}
  \label{fig:parameter-budget}
\end{figure}

Figure~\ref{fig:parameter-budget} compares marginal and absolute-return-ACF
performance against the number of trainable generator coefficients. For $L$
layers, ZZ and time evolution have $L$ coefficients, while ZZ--CRX has $2L$;
under the imposed sharing rule, these counts are independent of the number of
qubits. At the evaluated depths $L=1,\ldots,4$, the problem-informed families
therefore occupy the 1--8 coefficient region, compared with 48--120
coefficients for HEA and 29--257 for TCNN. Within this region, they retain
meaningful marginal and temporal performance without collapsing to the poor
marginal fit of the simplest control, although the more flexible benchmarks
still attain the lowest marginal distances. These comparisons quantify
parameter economy: the coefficient count measures the dimension of the
optimised generator parameterisation, rather than total computational or
hardware cost. Architecture and logical-operation counts are provided in
Online Resource~1, Table~S4.

The points within the problem-informed families also show that adding
coefficients does not lead monotonically to better performance. For ZZ--CRX,
the representative marginal $W_1$ worsens from 0.00249 at one layer (two
coefficients) to 0.00612 at four layers (eight coefficients). Time evolution
shows the opposite marginal trend, improving from 0.00227 to 0.00176 between
one and four layers, but its absolute-return ACF error increases from 0.07291 to
0.10161. The two architectures therefore show different depth trends. Because
architecture, depth, and optimisation change together, these points do not
identify parameter count as the cause.

The matched two-coefficient case makes the role of architecture particularly
clear. ZZ--CRX layer 1 outperforms the two-parameter classical control on
both objectives retained in the main figure, showing that how the generator is
organised matters alongside the number of optimised coefficients.
Online Resource~1, Figure~S12, completes
this comparison with the other temporal diagnostics: the control remains
relatively poor for raw-return ACF fidelity but performs comparatively well for
leverage, and is therefore not uniformly inferior.

\begin{table*}[t]
  \centering
  \caption{Representative performance. Configurations and runs are selected using marginal $W_1$
  following the procedure in Section~\ref{sec:model-selection-evaluation}.
  Temporal errors are mean absolute differences between empirical and generated
  median window-level curves over lags 1--4. Lower values are better for all
  metrics.}
  \label{tab:representative-performance}
  \small
  \setlength{\tabcolsep}{3.6pt}
  \begin{tabular}{@{}lrrrrrrr@{}}
    \toprule
    Generator & Active & Run & Marginal & Q--Q & $|r|$ ACF & Raw ACF & Leverage \\
              & params. &     & $W_1$    & RMSE  & MAE       & MAE     & MAE \\
    \midrule
    ZZ, $L=3$ & 3 & 2 & 0.00185 & 0.00323 & 0.04286 & 0.03448 & 0.01492 \\
    ZZ--CRX, $L=1$ & 2 & 1 & 0.00249 & 0.00466 & 0.05967 & 0.02130 & 0.02412 \\
    Time evolution, $L=4$ & 4 & 0 & 0.00176 & 0.00256 & 0.10161 & 0.01859 & 0.01099 \\
    \midrule
    HEA, $L=2$ & 72 & 2 & 0.00106 & 0.00256 & 0.14122 & 0.01553 & 0.03635 \\
    TCNN, $W=6$ & 157 & 2 & 0.00093 & 0.00218 & 0.08408 & 0.01507 & 0.00864 \\
    Two-parameter classical & 2 & 1 & 0.00607 & 0.01247 & 0.07920 & 0.05984 & 0.01505 \\
    \bottomrule
  \end{tabular}
  \vspace{2pt}

  \begin{minipage}{\textwidth}
    \footnotesize The first three rows are the proposed problem-informed
    generators; the remaining rows are reference models.
  \end{minipage}
\end{table*}

\newpage

\section{Discussion}

\subsection{Structural constraints and interpretability}

The problem-informed generators make a deliberate trade: they give up
unrestricted variational flexibility for a smaller and more inspectable model
family. Each operator enters with a hypothesis about the statistic it may
influence, while coefficient sharing restricts the number of independent
directions explored during training. The resulting generators occupy a much
smaller active-coefficient regime than the hardware-efficient benchmark and
make selected one- and two-parameter landscapes directly visualisable. Their
main contribution therefore lies not only in the fidelity of the generated
returns, but also in the traceable connection between circuit construction and
observable statistical responses.

We discuss these three aspects of interpretability in relation to the results.
At the design stage, each operator family enters the candidate pool with an
explicit hypothesis about its role. This resembles the use of chemically
motivated excitation operators in unitary coupled cluster (UCC) ansätze
\citep{romero2018strategies} and in operator pools for the Adaptive
Derivative-Assembled Pseudo-Trotter ansatz variational quantum eigensolver
(ADAPT-VQE) \citep{grimsley2019adaptive}. In ADAPT-VQE, operators are selected
iteratively during optimisation to grow the ansatz
\citep{grimsley2019adaptive}. Here, the operators are
phenomenological encodings rather than terms derived from a microscopic market
Hamiltonian. The one-layer configurations show that a compact operator pool can
achieve competitive distributional fidelity within a restricted variational
space.

The second stage is empirical: the intended operator role is compared with a
measurable statistical response. ZZ and ZZ--CRX provide an example. Setting all
directed-control coefficients to zero recovers the ZZ architecture. Among
independently trained and selected configurations, ZZ--CRX exhibits
predominantly lower median leverage than ZZ at matched depths, consistent with
the intended directional hypothesis.

The third stage is landscape-level interpretability. Coefficient sharing makes
the selected two-parameter loss surfaces directly inspectable. Their contours
show which parameter directions most strongly affect the sampled losses and
where changing one coefficient reinforces or offsets the effect of another.
Architecture-specific contour structures provide additional insight into how
coefficients accumulate or interact. The landscapes also reveal where
parameter changes favour marginal and temporal diagnostics differently.
In the corrected full-period ZZ--CRX scan, the newly exposed temporal basin
does not coincide with the marginally selected checkpoint region; the marker
placement therefore reflects the stated marginal-only selection rule rather
than evidence that training found a joint compromise.
Together, these stages connect an operator-level hypothesis to an empirical
response and then to inspectable parameter contributions and interactions.

The same compactness also defines the main risk of the approach. Because the
mechanism generating a financial distribution is unknown, an incomplete
operator pool may exclude useful variational directions. The digitised ans\"atze
apply noncommuting operators in a fixed sequence, allowing their effects to
reinforce, suppress, or redirect one another in an order-dependent manner. The
zero-coefficient nesting relation between ZZ--CRX and ZZ therefore does not
turn the reported comparison of independently trained and selected
configurations into a controlled intervention: the other fitted coefficients
were not held fixed. This interplay makes the contribution of each operator to
the final observable response harder to disentangle. Causal attribution to
individual operator terms requires controlled ablations.

\subsection{Limitations and outlook}

Three directions follow from the present limitations.

First, the study uses a particular Bitcoin sample partitioned into overlapping 16-step windows with unit stride. Increasing the window length to 30–40 steps would provide more within-window pairs at a given lag, potentially reduce the sampling variability of autocorrelation and leverage estimates, and permit evaluation at longer lags. A longer Bitcoin series would help maintain a comparable number of training windows. The larger register would substantially increase the cost of full-state simulation and training, motivating a move to quantum hardware for this extension. Several architectural features support this route: single-layer configurations already show promising marginal fidelity and selected short-range temporal behaviour, while one-dimensional loop connectivity accommodates the nearest-neighbour interactions and boundary connection without routing SWAPs on hardware with matching connectivity. Increasing output length could therefore increase circuit width and gate count while retaining a single ansatz layer. This avoids additional layer repetitions, although the compiled gate depth can still grow with register size, particularly for the ordered controlled-rotation sweeps. The resulting circuit depth and exposure to decoherence would therefore require hardware-specific assessment.

Second, the dual-observable readout requires basis-dual operators and a dedicated boundary construction, which complicate operator-level interpretation. A useful extension is a $Z$-only problem-informed generator, which would make the connection between operator choices and observable statistical responses more direct. Under this reformulation, however, a direct analogue of the time-evolution architecture studied here would be observationally trivial: its post-encoding $ZZ$ interactions commute with every measured $Z_i$ and therefore cannot alter the $Z$-basis readout. The operator pool would consequently require noncommuting problem-informed interactions, such as the controlled-$X$ terms used in the ZZ--CRX architecture, to ensure that the trainable interaction coefficients produce measurable responses.

Third, the present dataset exhibits weak raw-return autocorrelation and an uncertain leverage signal, limiting its ability to distinguish some aspects of temporal fidelity. Further experiments should examine assets and market regimes with more pronounced volatility clustering and leverage. These would provide clearer tests of the intended statistical roles of the problem-informed operators and assess whether the design protocol transfers beyond the present Bitcoin sample.

\section{Conclusion}

We developed and evaluated problem-informed quantum generators for short
financial return windows. The eight-qubit circuits used 1--8 trainable
coefficients, recovered broad features of the Bitcoin return distribution, and,
in selected configurations, retained short-range absolute-return dependence.
Shallow configurations remained competitive within the proposed families,
while neither added depth nor larger coefficient counts produced consistent
improvements. The TCNN reference achieved stronger marginal fidelity, whereas
selected ZZ and ZZ--CRX configurations better retained short-range
absolute-return dependence. In the matched two-coefficient comparison, ZZ--CRX
outperformed the classical control on the two displayed objectives, while the
control performed better on selected remaining temporal diagnostics.

The interpretability contribution combines explicit operator hypotheses with
measured statistical responses and inspectable parameter landscapes. Across
matched depths, independently trained ZZ--CRX configurations exhibited
predominantly lower median leverage than ZZ, consistent with the intended
directional modulation. This shift did not imply uniformly better agreement
with empirical leverage, whose weak signal limits the assessment of fidelity.
The low-dimensional landscapes further showed that marginal and
absolute-return-dependence diagnostics often favoured different parameter
regions. Expanding the ZZ--CRX directed coefficient to its full period exposed
a lower temporal basin without bringing the marginally selected checkpoints
into that region. These results make the statistical consequences and competing
preferences of compact circuit designs directly examinable. A central next
step is to refine the operator pools to improve marginal and temporal agreement
more consistently while preserving shallow circuits and interpretable
parameter roles.

\backmatter
\section*{Statements and declarations}

\paragraph{Funding.}
The authors declare that no funds, grants, or other support were received during
the preparation of this manuscript.

\paragraph{Competing interests.}
The authors have no relevant financial or non-financial interests to disclose.

\paragraph{Author contributions.}
Conceptualization: Huijie Guan, Shengbin Wang, Wenhao Chu, and Hantao Sun;
supervision and project administration: Wenhao Chu and Hantao Sun;
methodology: Huijie Guan, Jiale Zhang, Xuanhao Wang, and Shengbin Wang;
software and data curation: Jianjian Gao, Dongdong Li, and Jiale Zhang;
investigation, formal analysis, validation, and visualization: Huijie Guan,
Jiale Zhang, Shuxin Xie, and Jian Ding; writing---original draft: Huijie Guan
and Xuanhao Wang; writing---review and editing: all authors. All authors read
and approved the final manuscript.

\paragraph{Data availability.}
The processed daily Bitcoin dataset used in this study was derived from
\emph{Bitcoin Historical Data}, published by Zielak (mczielinski) on Kaggle
\citep{zielak2026bitcoin}. The original minute-level file,
\texttt{btcusd\_1-min\_data.csv}, is not redistributed and can be obtained from
the cited source under the CC BY-SA 4.0 licence. The processed dataset,
generated evaluation windows, selected trained parameters, analysis tables,
and provenance manifests are not included with this arXiv submission.

\paragraph{Code availability.}
The preprocessing, model-training, evaluation, and figure-generation code is
not included with this arXiv submission.

\paragraph{Declaration of generative AI use.}
During preparation of this manuscript, the authors used OpenAI Codex to assist with manuscript organisation, preliminary drafting, language revision, and code review. The authors reviewed and edited all AI-assisted material and take full responsibility for the content of the article.

\bibliography{references}

\end{document}